\documentclass[prx, aps, twocolumn]{revtex4-1}
\usepackage{amsmath,amssymb}
\usepackage{enumitem}
\usepackage{graphicx}
\usepackage{tikz}
\usetikzlibrary{positioning}
\usepackage{color}
\usepackage{siunitx}
\usepackage{gensymb}

\usepackage{lineno}
\newcommand*\patchAmsMathEnvironmentForLineno[1]{%
  \expandafter\let\csname old#1\expandafter\endcsname\csname #1\endcsname
  \expandafter\let\csname oldend#1\expandafter\endcsname\csname end#1\endcsname
  \renewenvironment{#1}%
     {\linenomath\csname old#1\endcsname}%
     {\csname oldend#1\endcsname\endlinenomath}}%
\newcommand*\patchBothAmsMathEnvironmentsForLineno[1]{%
  \patchAmsMathEnvironmentForLineno{#1}%
  \patchAmsMathEnvironmentForLineno{#1*}}%
\AtBeginDocument{%
\patchBothAmsMathEnvironmentsForLineno{equation}%
\patchBothAmsMathEnvironmentsForLineno{align}%
\patchBothAmsMathEnvironmentsForLineno{flalign}%
\patchBothAmsMathEnvironmentsForLineno{alignat}%
\patchBothAmsMathEnvironmentsForLineno{gather}%
\patchBothAmsMathEnvironmentsForLineno{multline}%
}

\makeatletter
\def\p@paragraph{\thesection-}
\makeatother

\newcommand{\target}{^*}

\newcommand{\Dt    }{\Delta t}
\newcommand{\Dm    }{\Delta m}
\newcommand{\Dalpha}{\Delta\alpha}
\newcommand{\pER}{p_\mathrm{con}}
\newcommand{\tauhp}{\tau_\mathrm{hp}}
\newcommand{\tauprime}{\tau^{\prime}}

\newcommand{\taubp}{\tau}

\newcommand{\avg}[1]{\overline{#1}}
\newcommand{\expectation}[1]{\langle #1\rangle}

\newcommand{\invitro}{{\it in vitro}}
\newcommand{\Invitro}{{\it In vitro}}
\newcommand{\invivo}{{\it in vivo}}
\newcommand{\Invivo}{{\it In vivo}}

\begin{document}
\title{Homeostatic plasticity and external input shape neural network dynamics}
\author{Johannes Zierenberg$^{1,2}$}
\thanks{these authors contributed equally}
\author{Jens Wilting$^{1}$}
\thanks{these authors contributed equally}
\author{Viola Priesemann$^{1,2}$}
\affiliation{
  \mbox{$^1$ Max Planck Institute for Dynamics and Self-Organization, Am Fassberg 17, 37077 G{\"o}ttingen, Germany,}\\
  \mbox{$^2$ Bernstein Center for Computational Neuroscience, Am Fassberg 17, 37077 G{\"o}ttingen, Germany,}
}

\begin{abstract}
  \Invitro{} and \invivo{} spiking activity clearly differ. Whereas networks
  \invitro{} develop strong bursts separated by periods of very little spiking
  activity, \invivo{} cortical networks show continuous activity.  This is
  puzzling considering that both networks presumably share similar single-neuron
  dynamics and plasticity rules. We propose that the defining difference between
  \invitro{} and \invivo{} dynamics is the strength of external input.
  \Invitro{}, networks are virtually isolated, whereas \invivo{} every brain
  area receives continuous input. We analyze a model of spiking neurons
  in which the input strength, mediated by spike rate homeostasis, determines
  the characteristics of the dynamical state. In more detail, our analytical
  and numerical results on various network topologies show consistently that
  under increasing input, homeostatic plasticity generates distinct dynamic
  states, from bursting, to close-to-critical, reverberating and irregular
  states. This implies that the dynamic state of a neural network is not fixed
  but can readily adapt to the input strengths. Indeed, our results match
  experimental spike recordings \invitro{} and \invivo{}: the \invitro{}
  bursting behavior is consistent with a state generated by very low network
  input ($<0.1\%$), whereas \invivo{} activity suggests that on the order of
  $1\%$ recorded spikes are input-driven, resulting in reverberating dynamics.
  Importantly, this predicts that one can abolish the ubiquitous bursts of
  \invitro{} preparations, and instead impose dynamics comparable to \invivo{}
  activity by exposing the system to weak long-term stimulation, thereby
  opening new paths to establish an \invivo{}-like assay \invitro{} for basic
  as well as neurological studies.
\end{abstract}

\maketitle

\section{Introduction}
Collective spiking activity clearly differs between \invitro{} cultures and
\invivo{} cortical networks (see examples in Fig.~\ref{figExperiment}). Cultures
\invitro{} typically exhibit stretches of very little spiking activity,
interrupted by strong bursts of highly synchronized or coherent
activity~\cite{robinson1993,pelt2004, chiappalone2006, wagenaar2006,
orlandi2013, vardi2016, beggs2003}. In contrast, spiking activity recorded from
cortex in awake animals \invivo{} lacks such pauses, and instead shows
continuous, fluctuating activity. These fluctuations show a dominant
autocorrelation time that was proposed to increase hierarchically across
cerebral cortex, from sensory to frontal areas~\cite{murray2014}. Moreover,
depending on experimental details such as brain area, species and vigilance
state, one also observes evidence for asynchronous-irregular (AI)
dynamics~\cite{burns1976,softky1993}, oscillations~\cite{gray1989,gray1994,
buzsaki2004}, or strong fluctuations associated with criticality, bistability or
up-and-down states~\cite{breakspear2017, priesemann2009, priesemann2013,
bellay2015, wilson2008, stern1997, cossart2003}. These states differ not only
in strength and structure of fluctuations, but also in synchrony among neurons,
from uncorrelated  to fully synchronized spiking.  The observation of such a
vast range of dynamic states is puzzling, considering that the dynamics of all
networks presumably originate from similar single-neuron physiology and
plasticity mechanisms.

One particular plasticity mechanism that regulates neural activity on a long
time scale is homeostatic plasticity~\cite{turrigiano1998, lissen1998,
obrien1998, turrigiano2004, davis2006, marder2013}. Homeostatic plasticity can
be implemented by a number of physiological candidate mechanisms, such as
redistribution of synaptic efficacy~\cite{markram1996, tsodyks1997}, synaptic
scaling~\cite{turrigiano1998, lissen1998,obrien1998,fong2015}, adaptation of
membrane excitability~\cite{davis2006,pozo2010}, or through interactions with
glial cells~\cite{pitta2016, virkar2016}. All these mechanisms have in common
that they implement a slow negative feedback-loop in order to maintain a
certain target spike rate and stabilize network dynamics. In general, they
reduce (increase) excitatory synaptic strength or neural excitability if the
spike rate is above (below) a target rate, allowing compensation against a
potentially unconstrained positive feedback-loop through  Hebbian-type
plasticity~\cite{bienenstock1982, miller1994,
abbott2000,turrigiano2000,zenke2013,keck2017,zenke2017}. Recent results
highlight the involvement of homeostatic plasticity in generating robust yet complex activity
dynamics in recurrent networks~\cite{naude2013,hellyer2016, gjorgjieva2016}.

To understand the physiological mechanisms behind this large set of dynamic
states, different model networks have been proposed that reproduce one or a set
of states. To name a few examples, deafferentiation in combination with
homeostatic scaling can generate bursts~\cite{froehlich2008a}; 
the interplay between excitation and inhibition may lead to oscillations, 
synchronous-regular activity, or asynchronous-irregular
activity~\cite{wilson1972,vogels2005,fries2007,brunel2000}, where switching
between dynamic states can be induced by varying the input~\cite{brunel2000,
lercher2015, munoz2018}; synaptic facilitation and depression promote regular
and irregular network dynamics~\cite{tsodyks1998, levina2007, levina2009};
plasticity at inhibitory synapses can stabilize irregular
dynamics~\cite{vogels2011, effenberger2015}, whereas specific types of
structural~\cite{bornholdt2000, tetzlaff2010, kossio2018} or synaptic~\cite{levina2007,
levina2009, arcangelis2006, bonachela2010, costa2015, kessenich2016, campos2017,
herrmann2017, millman2010, munoz2017,delPapa2017} plasticity foster strong
temporal fluctuations characteristic for a critical state; last but not least,
homeostasis is necessary to achieve stable dynamics in recurrent networks with
spike-timing dependent plasticity (STDP) or Hebbian-type synaptic plasticity
(e.g. Refs.~\cite{lazar2009, litwinkumar2014, zenke2015, miner2016, tosi2017,
keck2017}).  
Overall, the dynamic state depends on all aspects: single-neuron properties,
synaptic mechanisms, network topology,  plasticity rules, and input
characteristics.
Recalling that the single-neuron properties, synaptic mechanisms, as well as
plasticity rules are presumably very similar \invitro{} and \invivo{}, these
factors are unlikely to explain the observed differences.


In this study, we propose that the input strength is the defining difference
between \invitro{} and \invivo{} dynamics. \Invitro{} systems are completely
isolated, whereas \invivo{} networks receive continuous input from sensory
modalities and other brain areas. Under these different conditions, we propose
that homeostatic plasticity is a sufficient mechanism to promote
self-organization to a diverse set of dynamic states by mediating the
interplay between external input rate and neural target spike rate. Treating the
external input as a control parameter in our theoretical framework, allows us to
alter the network dynamics from bursting, to fluctuating, to irregular. Thereby,
our framework offers testable predictions for the emergence of characteristic
but distinct network activity \invitro{} and \invivo{}.

Based on our theory, we derive explicit experimental predictions and
implications: (1) The direct relation between dynamic state, spike rate and
input rate enables us to quantify the amount of input the neural network
receives, e.g., in mildly anesthetized cat V1, we estimate an input rate of
$\mathcal{O}(\SI{0.01}{Hz/neuron})$.
(2) This implies that about $2\%$ of cortical activity in cat V1 are imposed by
the input, whereas $98\%$ are generated by recurrent activation from within the
network.
(3) Our results suggest that one can alter the dynamic state of an experimental
preparation by altering the input strength. Importantly, we predict for
\invitro{} cultures that increasing the input rate to about
$\mathcal{O}(\SI{0.01}{Hz/neuron})$ would be sufficient to induce
\invivo{}-like dynamics.

\section{Experimental observations}
\label{secExperiment}
\begin{figure*}[]
  \includegraphics[]{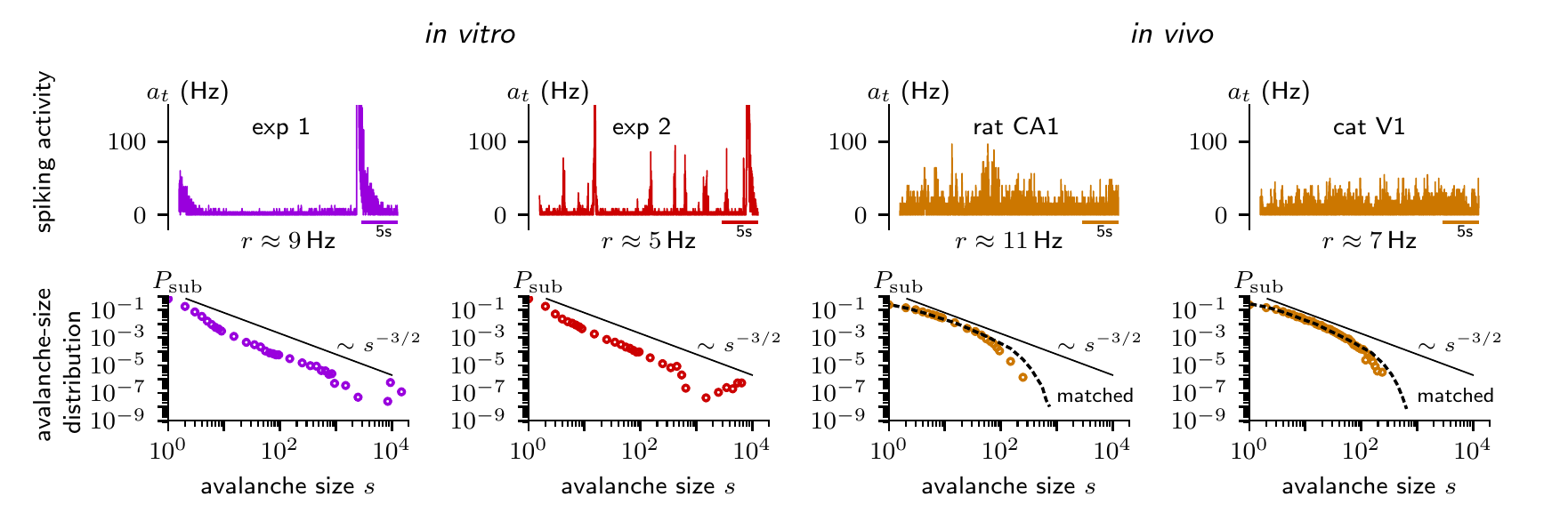}
  \caption{%
    Examples of dynamic states observed in experiments. \Invitro{} spike
    recordings are from cultures of dissociated rat cortical
    neurons~\cite{wagenaar2006}. \Invivo{} recordings are from right dorsal
    hippocampus in an awake rat during an open field task~\cite{mizuseki2009a}
    and from primary visual cortex in a mildly anesthetized
    cat~\cite{blanche2009}. Top row shows population spiking activity
    (Appendix~\ref{secAppendixNeuralActivity}) from
    30--60 single or multi-units ($\Dt=\SI{4}{ms}$) with average neural firing
    rate $r$; bottom row shows subsampled avalanche-size distributions
    (Appendix~\ref{secAppendixAvalanche}).  
    Solid lines indicate the power-law behavior $s^{-3/2}$ expected for a
    critical branching process. 
    Dashed lines correspond to distributions obtained from branching networks
    matched to the experiments of 
       rat CA1 ($\tau=\SI{2}{s}$, $r\target=\SI{11}{Hz}$, $h=\SI{5.5e-3}{Hz}$, $n=31$, $N=10^4$, $\Dt=\SI{1}{ms}$, $\tauhp=\SI{e5}{s}$)
    and cat V1 ($\tau=\SI{0.2}{s}$, $r\target=\SI{ 7}{Hz}$, $h=\SI{3.5e-2}{Hz}$, $n=50$, $N=10^4$, $\Dt=\SI{1}{ms}$, $\tauhp=\SI{e5}{s}$).
    For details and a definition of parameters see
    Appendix~\ref{secAppendixMatchExp}. 
    \label{figExperiment}
  }
\end{figure*}
To demonstrate characteristic neural activity \invitro{} and \invivo{}, we
analyzed exemplary recordings of spiking activity. Data sets included cultures
of dissociated cortical neurons~\cite{wagenaar2006, wagenaarURL}, as well as
hippocampus of foraging rats~\cite{mizuseki2009a, mizuseki2009b} and
visual cortex of mildly anesthetized cats~\cite{blanche2006, blanche2009} (see
Appendix~\ref{secAppendixExperiments} for details).
Note that all preparations were inevitably subsampled, as spikes were recorded
only from a small number of all neurons.
For illustrative purposes, we focus on the average (subsampled) spiking activity
$a_t$ and the (subsampled) avalanche-size distribution $P_\mathrm{sub}$
(see Appendix~\ref{secAppendixNumerics} for details).

The spiking activity \invitro{} shows bursting behavior
(Fig.~\ref{figExperiment}), i.e., stretches of very low activity interrupted by
periods of synchronized activity. The subsampled avalanche-size distributions
$P_{\rm sub}(s)$ exhibits partial power-law behavior resembling $P(s)\sim
s^{-3/2}$ as expected from a critical branching process~\cite{harris1963}, and
conjectured for the synchronous-irregular regime~\cite{touboule2017}. However,
in addition $P_{\rm sub}(s)$ also shows a peak at large avalanche sizes, which
may indicate either finite-size effects, supercriticality, or characteristic
bursts~\cite{levina2017}. 

In contrast, the spiking activity \invivo{} shows fluctuating dynamics
(Fig.~\ref{figExperiment}). These have been described as reverberating dynamics,
a dynamic state between critical and irregular dynamics~\cite{wilting2018},
characterized by a finite autocorrelation time of a few hundred milliseconds.
The subsampled avalanche-size distributions $P_{\rm sub}(s)$ can be
approximated by a power-law for small $s$ but show a clear exponential tail.
The tails indicate slightly subcritical dynamics~\cite{priesemann2014},
especially because deviations in the tails are amplified under
subsampling~\cite{priesemann2009, priesemann2013, levina2017}. 

In sum, the spiking activity and the corresponding avalanche-size distributions
clearly differ between \invitro{} and \invivo{} recordings. Remarkably,
however, the average neural firing rate $r$ is similar across the different
experimental setups. 

\section{Model}
\label{secModel}
To investigate the differences between \invitro{} and \invivo{}, we make use of
a branching network, which approximates properties of neural activity
propagation. We extend the branching network by a negative feedback, which
approximates homeostatic plasticity. 

\subsection{Branching network}
In the brain, neurons communicate by sending spikes.
The receiving neuron integrates its input, and if the membrane potential
crosses a certain threshold, this neuron fires a spike itself. 
As long as a neuron does not fire, its time-varying membrane potential can be
considered to fluctuate around some resting potential.
In the following, we approximate the complex time-resolved process of action
potential generation and transmission in a stochastic neural model with
probabilistic activation. 

Consider a network of size $N$. Each node corresponds to an excitatory neuron,
and spike propagation is approximated as a stochastic process at discrete time
steps $\Dt$. If a neuron, described by the state variable $s_{i,t}\in\{0,1\}$,
is activated, it spikes ($s_{i,t}=1$), and immediately returns to its resting
state ($s_{i,t+1}=0$) in the next time step, unless activated again.
Furthermore, it may activate post-synaptic neurons $j$ with probability
$p_{ij,t}=w_{ij}\alpha_{j,t}$, where $w_{ij}\in\{0,1\}$ indicates whether two
neurons are synaptically connected, and $\alpha_{j,t}$ is a homeostatic scaling
factor.
%
%
In addition, each neuron receives network-independent external input at rate
$h_i$, representing external input from other brain areas, external stimuli, and
importantly also spontaneous spiking of single neurons generated independently
of pre-synaptic spikes (e.g. by spontaneous synaptic vesicle
release~\cite{kavalali2014,lenz2015}). The uncorrelated external input
homogeneously affects the network at rate $h_i=h$, modeled as Poisson processes
with an activation probability $1-e^{-h\Dt}\simeq h\Dt$.

This model can be treated in the framework of a branching
process~\cite{harris1963}, a linear process with a characteristic
autocorrelation time $\tau$ (see below). The population activity is
characterized by the total number of spiking neurons, $A_t=\sum_{i=1}^N
s_{i,t}$.  Each spike at time $t$ generates on average $m$ postsynaptic spikes
at time $t+1$ such that on average $\mathbb{E}(A_{t+1}|A_t) = m A_t$, where $m$
is called \textit{branching parameter}. The branching parameter can be defined
for each neuron individually: neuron $i$ activates on average
\begin{equation}\label{eqBPLocal}
  m_{i,t}=\sum_{j=1}^N w_{ij}\,\alpha_{j,t}
\end{equation}
of its post-synaptic neurons~\cite{haldeman2005}. This local branching parameter
$m_{i,t}$ thus quantifies the impact of a spike in neuron $i$ on the network.
The network average (denoted in the following with a bar) of $m_{i,t}$ generates
the (time-dependent) network branching parameter~\cite{millman2010}
\begin{equation}\label{eqBPNetwork}
  \avg{m}_t = \frac{1}{N}\sum_{i=1}^N m_{i,t}.
\end{equation}
 
The external input generates on average $Nh\Dt$ additional spikes per time step,
resulting in a driven branching process~\cite{heathcote1965, pakes1971}.  The
expected activity at time $t+1$ is then 
$\mathbb{E}(A_{t+1}|A_t) = m A_t + Nh\Dt$.
%
For $m<1$ the process is called \textit{subcritical}, meaning that individual
cascades of events will eventually die out over time. In this case, the temporal
average (denoted in the following as $\expectation{\cdot}$) of network activity
$A_t$ converges to a stationary distribution with average activity
\begin{equation}\label{eqBranchingProcessActivity}
  \expectation{A} = \frac{1}{T}\sum_{t=1}^T A_t
  \underset{T\rightarrow\infty}{\longrightarrow} \frac{Nh\Dt}{1-m}.
\end{equation}
Considering a homogeneous neural spike rate $r_i=r=\expectation{A}/N\Dt$ this implies 
\begin{equation}\label{eqBranchingProcessRate}
  r = \frac{h}{1-m}.
\end{equation}
A constant mean spike rate $r$, which can be considered a biological
constraint, is thus realized by adjusting either $m\in[0,1)$ or
$h\in[0,\infty)$.

The subcritical branching process ($m<1$) is stationary with the autocorrelation
function $C(l)=m^l$. The autocorrelation time can be identified by comparing
with an exponential decay $C(l)=e^{-l\Dt/\taubp}$, yielding~\cite{wilting2018}
\begin{equation}\label{eqTau}
  \taubp=-\Dt/\ln(m),
\end{equation}
which diverges as $m\to1$. 
At this divergence ($m=1$) the process is \textit{critical} and the activity
$A_t$ grows linearly in time with rate $h$. At criticality, assuming $h
\rightarrow 0$, the number of events $s$ in an avalanche triggered by a single
seed event, is distributed according to a power law $P(s)\sim
s^{-3/2}$~\cite{harris1963}.
For a non-vanishing $h$ in the subcritical regime ($m<1$), the avalanche-size
distributions show a rapid decay, if they can be measured at all under
persistent activity (Appendix~\ref{secAppendixAvalanche}).
Finally, for $m>1$, the process is called \textit{supercritical} and $A_t$ can
in principle grow to infinity. For a finite network, this of course is not
possible and will manifest in a peak of the avalanche-size distribution at large
avalanche sizes.  

For the computational model, we consider a network of $N=10^4$ neurons, which
represents the size of \invitro{} cultures and \invivo{} cortical hypercolumns.
The time step of the branching process has to reflect the causal signal
propagation of the underlying physiological network. Since realistic propagation
times of action potentials from one neuron to the next range from
$\SIrange{1}{4}{ms}$, we choose $\Dt=\SI{1}{ms}$. 
%
%
We consider three network topologies:

\begin{figure}[]
  \begin{tikzpicture}
    \node (A) at (-2.1cm,0) {\includegraphics[width=0.48\columnwidth]{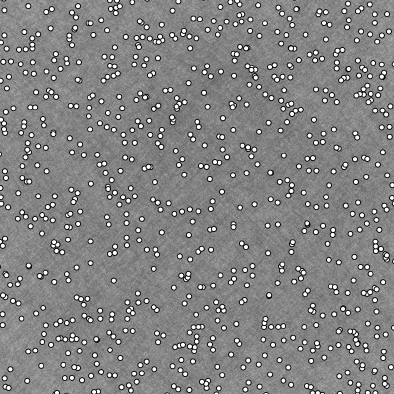}};
    \node (B) at ( 2.1cm,0) {\includegraphics[width=0.48\columnwidth]{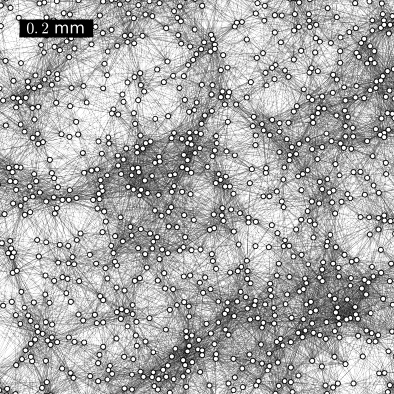}};
    \node at (-2.1cm,-2.3cm) {\bf a};
    \node at ( 2.1cm,-2.3cm) {\bf b};
  \end{tikzpicture}
  \caption{%
    Cutouts of two random network topologies. (\textbf{a}) Subset of randomly
    spaced nodes in an Erd\H{o}s-R{\'e}nyi (ER) network with $p=10^{-3}$. Note
    that connections cross the window also from neurons outside of the field of
    view, such that single connections cannot be distinguished visually in the
    sketch. (\textbf{b}) $1.4\times1.4$\SI{}{mm}$^2$ subset of spatially-clustered (SC) topology generated by axonal-growth
    rules~\cite{orlandi2013,hernandez2017}.
    \label{figNetworks}
  }
\end{figure}
\paragraph{Directed Erd\H{o}s-R{\'e}nyi (ER) network:}
As a standard model of network topology, we consider a network with random
directed connections. Each connection $w_{ij}=1$ is added with probability
$\pER$, excluding self-connections $(i,i)$. 
Then, the degree distribution of outgoing as well as incoming connections
follows a binomial distribution
%
%
with average degree $\avg{k} = \pER (N-1)\simeq\pER N$. We require
$\pER>\ln(N)/N$ to ensure that the graph is connected~\cite{erdoes1960}. The
connectivity matrix $w_{ij}$ is fixed throughout each simulation, such that
averaging over simulations with different network realizations results in a
quenched average. A cutout from an example graph is shown in
Fig.~\ref{figNetworks}\textbf{a}.

\paragraph{Spatially-clustered (SC) network:}
In order to consider a more detailed topology with dominant short-range
connections, we follow Orlandi \textit{et al.} who developed a model based on
experimental observations of \invitro{} cultures~\cite{orlandi2013,
hernandez2017}. Neural somata are randomly placed as hard discs with radius
$R_s=7.5\,\mu\text{m}$, to account for the minimal distance between cell
centers, on a $5\times5\,\text{mm}^2$ field.  From each soma an axon grows into
a random direction with a final length $l$ given by a Rayleigh distribution
$p(l)=(l/\sigma_l^2)~\exp(-l^2/2\sigma_l^2)$ with $\sigma_l=900\,\mu\text{m}$ and
average axonal length $\bar{l}\simeq1.1\text{mm}$.  The axonal growth is a
semiflexible path with segments of size $\Delta l=10\,\mu\text{m}$ and
orientation drawn from a Gaussian distribution relative to the previous segment
with $\sigma_\theta=15\,\degree$. A connection with another neuron is formed
with probability $1/2$ if the presynaptic axon intersects with the dendritic
tree of a postsynaptic neuron~\cite{wen2009}. The dendritic tree is modeled as a
disc around the postsynaptic neuron with radius $R_d$ drawn from a Gaussian
distribution with mean $\bar{R}_d=300\,\mu\text{m}$ and
$\sigma_d=20\mu\text{m}$. A cutout from an example graph is shown in
Fig.~\ref{figNetworks}\textbf{b}.

\paragraph{Annealed-average (AA) network:}
We consider in addition a network with $k$ dynamically changing random
connections (annealed average). The connections are distinguishable, exclude
self-connections, and are redrawn every time step. This model approximates the
otherwise numerically expensive fully connected network (ER with $\pER=1$) with
a global $m_t$ by choosing $\alpha_{j,t}=m_t/k$.  In practice, we chose $k=4$,
which produces analogous dynamics to the fully-connected ($\avg{k}\approx10^4$)
network as long as $m_t<4$. 

Error bars are obtained as statistical errors from the fluctuations between
independent simulations, which includes random network realizations
$\{w_{ij}\}$ for ER and SC.

\subsection{Homeostatic plasticity}
In our model, homeostatic plasticity is implemented as a negative feedback,
which alters the synaptic strength on the level of the post-synaptic neuron (the
homeostatic scaling factor $\alpha_{j,t}$) to reach a target neural firing rate
$r_j\target$. 
%
We consider a linear negative feedback with time constant $\tauhp$, which
depends solely on the (local) activity of the postsynaptic neuron $s_{j,t}$
\begin{equation}\label{eqHomeostaticPlasticityNeuron}
  \Dalpha_{j,t} =  (\Dt\,r_j\target - s_{j,t})\left(\frac{\Dt}{\tauhp}\right),
\end{equation}
i.e., adapting a neuron's synaptic strength does not rely on information about
the population activity $A_t$. Since $\alpha_{j,t}$ is a probability, we
additionally demand $\alpha_{j,t}\geq0$.
Equation \eqref{eqHomeostaticPlasticityNeuron}
considers homeostatic plasticity to directly couple to all postsynaptic synapses
of any given neuron $j$. This can be implemented biologically as autonomous synaptic
processes or somatic processes, such as translation and transcription. 
In order to further reduce complexity, we assume a uniform target rate
$r_j\target=r\target$, while in fact experiments show a broad (log-normal)
spike-rate distribution~\cite{buzsaki2014,hengen2016}. Preliminary tests for a
log-normal target rate distribution in ER networks ($\pER=0.1$) showed
consistent results.
In our simulations, we typically consider a biologically motivated target rate
$r\target=\SI{1}{Hz}$ and a homeostatic timescale of the order of an hour,
$\tauhp=\SI{e3}{s}$. 

\begin{figure*}[]
  \includegraphics{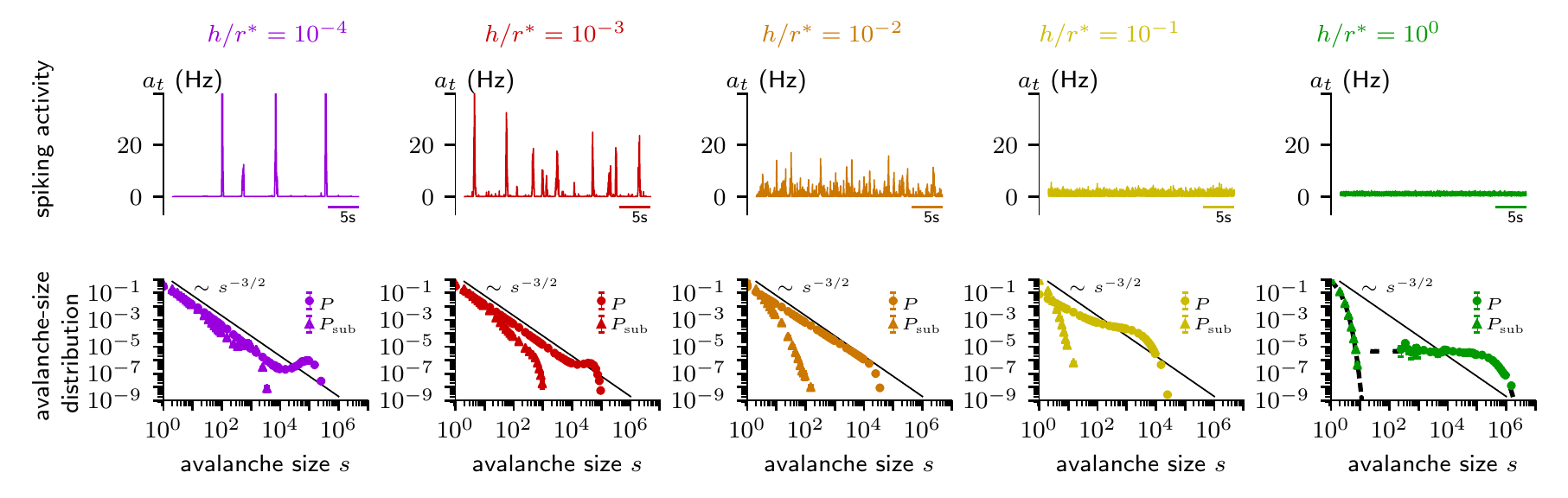}
  \caption{%
    Homeostatic plasticity induces diverse dynamic states by regulating
    recurrent network interactions, mediating input rate $h$ and target neural
    firing rate $r\target$. 
    %
    The annealed-average network reproduces bursting ($m>1$,
    $h/r\target\leq10^{-3}$, purple\--red), fluctuating ($m\approx
    0.99$,$h/r\target\approx10^{-2}$ and $m\approx0.9$,
    $h/r\target\approx10^{-1}$, orange\--yellow), and irregular ($m\approx0$,
    $h/r\target=1$, green) dynamics. 
    The top row shows examplary spiking activity $a_t=A_t/N\Dt$
    (Appendix~\ref{secAppendixNeuralActivity}); 
    the bottom row shows avalanche-size distributions $P(s)$ ($n=N$,
    circles) and subsampled avalanche-size distributions $P_\mathrm{sub}(s)$
    ($n=100$, triangles) averaged over 12 independent
    simulations (Appendix~\ref{secAppendixAvalanche}). 
    Solid lines show the power-law distribution $P(s)\propto
    s^{-3/2}$~\cite{harris1963}, dashed lines show the analytical avalanche-size
    distribution of a Poisson process~\cite{shriki2017}.  Parameters:  $N=10^4$,
    $\tauhp=\SI{e3}{s}$, $r^\ast=\SI{1}{Hz}$, $\Delta t=\SI{1}{ms}$.
    \label{figOverviewMF}
  }
\end{figure*}
%
\section{Results}
\label{secResults}

Including homeostatic plasticity in our model generates a broad range of dynamic
states, depending on the external input. Figure~\ref{figOverviewMF} shows
qualitatively representative results obtained for AA networks.
For strong input ($h=\mathcal{O}(r\target)$), the network organizes itself into
a dynamic state where neural firing is solely driven by the input fluctuations,
resembling an asynchronous-irregular state (green). Here, temporal and pairwise
spike count cross-correlations approach zero, and the avalanche-size
distribution matches the analytic solution for a Poisson
process~\cite{shriki2017} shown as dashed lines. 
For weaker input ($h<r\target$) the system tunes itself towards fluctuating
dynamics (orange\--yellow). The average neural rate and sub-sampled
avalanche-size distributions are qualitatively similar to reverberant \invivo{}
dynamics with autocorrelation times of several hundred milliseconds
(Fig.~\ref{figExperiment}). In this regime, the temporal correlations increase
when weakening the input, approaching close-to-critical dynamics, characterized
by a power-law distribution $P(s)=s^{-3/2}$~\cite{harris1963}, at the lower end
of the regime. 
Decreasing the input even further ($h\ll r\target$) leads to bursting behavior,
characterized by silent periods which are interrupted by strong bursts. These
bursts are apparent
as peak in the avalanche-size distribution at large avalanche sizes
(purple\--red). In this regime, the network steadily increases its synaptic
strengths during silent periods until a spike initiates a large burst, which in
turn decreases the synaptic strengths drastically, and so on (cf.
Appendix~\ref{secAppendixNonStat}). This regime captures the qualitative features
of bursting \invitro{} dynamics (Fig.~\ref{figExperiment}).

In the following, we derive a quantitative description of the three regimes
sketched above. 
To quantify the dynamic state, we consider the temporal average of the
branching parameter $m=\expectation{\avg{m}}$, as well as the associated
autocorrelation time $\tau$ of the population activity.

%
%

\begin{figure*}
  \begin{center}
  \begin{tikzpicture}
    \node (a) at (-4.5cm, 0cm) {\includegraphics{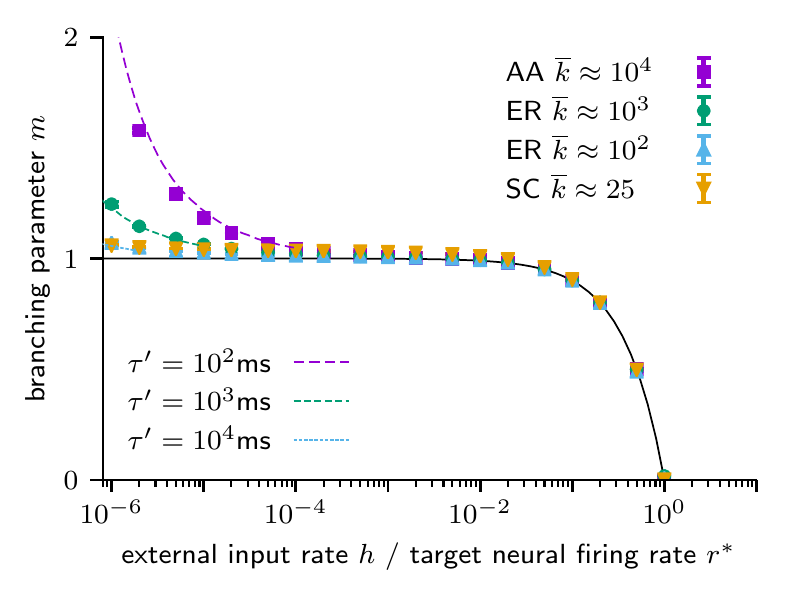}};
    \node (b) at (+4.5cm, 0cm) {\includegraphics{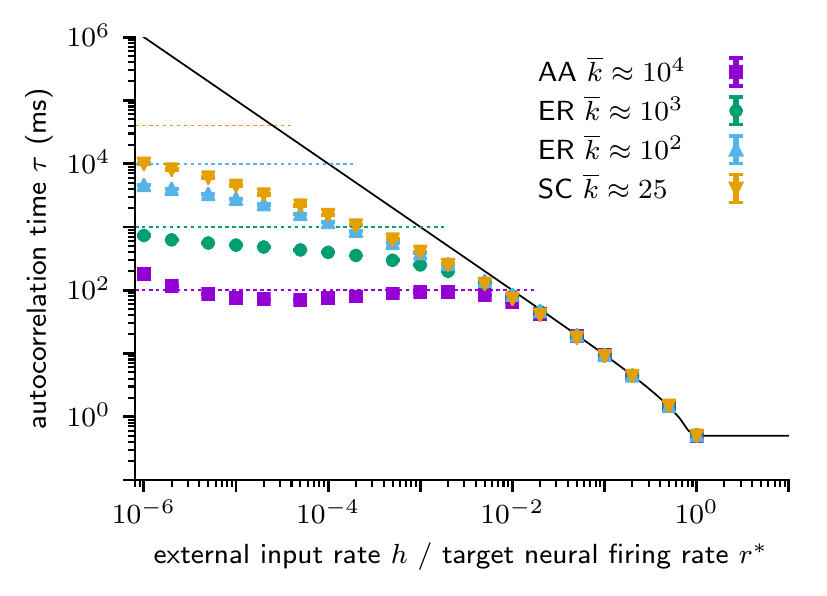}};
    \node [above left=-0.8cm of a] {\bf a};
    \node [above left=-0.8cm of b] {\bf b};
  \end{tikzpicture}
  \end{center}
  \caption{%
    Quantitative distinction between dynamic states induced in neural networks
    of different topologies by homeostatic plasticity as a function of
    normalized input rate $h/r\target$. Data points are averages over 12
    independent simulations ($N=10^4$, $\tau_\alpha=\SI{e3}{s}$, $r^
    \ast=\SI{1}{Hz}$, $\Delta t=\SI{e-3}{s}$) with connections generated
    according to annealed-average (AA), Erd\H{o}s-R{\'e}nyi (ER) or
    spatially-clustered (SC) topologies with average number of connections
    $\overline{k}$. Solid lines show the mean-field solution
    \eqref{eqMFprediction}, dashed lines represent (semi-analytical)
    approximations of the bursting regime. (\textbf{a}) Branching parameter
    $m=\expectation{\avg{m}}$ varies from irregular ($m\approx0$), to
    fluctuating ($m\lessapprox1$), to bursting ($m>1$) dynamics. The behavior in
    the bursting regime strongly depends on the network timescale
    \mbox{$\tauprime=\tauhp/\overline{k}$}.
    (\textbf{b}) Integrated autocorrelation time of the network
    population activity (Appendix~\ref{secAppendixTau}) shows a
    crossover from irregular [\mbox{$\tau=\mathcal{O}(\Delta t)$}], over
    fluctuating [$\tau=-\Delta t/\ln(1-h/r^\ast)$] to bursting
    ($\tau\approx\tauprime$) dynamics.
    \label{figHomPlastDynamicState}
  }
\end{figure*}
\subsection{Mean-field solution}
If we assume that $\tauhp$ is sufficiently large (i.e.  slow homeostatic
plasticity), then $\Delta\alpha_j\approx0$ and the dynamics of the network is
fully determined by the approximately constant branching parameter
$\avg{m}_t\approx m$. In this regime, \eqref{eqBranchingProcessRate} holds and
combined with \eqref{eqTau} and \eqref{eqHomeostaticPlasticityNeuron} we obtain
the mean-field solution
\begin{align}\label{eqMFprediction}
  m = 1 - h/r^\ast \quad\text{and}\quad \tau=-\Delta t/\ln(1-h/r^\ast).
\end{align}
Hence, with decreasing input rate $h$, recurrent network activation ($m$)
increases, i.e., perturbations cause a stronger network response and the
autocorrelation time increases (Fig.~\ref{figHomPlastDynamicState}, solid
lines). 
 
In the light of this mean-field solution, we discriminate the three
characteristic regimes as follows. 
First, we define the \textit{input-driven regime} by \mbox{$m\leq 0.5$ and
$\tau\approx\Delta t$.} Here, the network activity is dominated by input
($h=\mathcal{O}(r^\ast)$), and thus the dynamics follows the input statistics
and becomes irregular. Second, we define the \textit{fluctuating regime} for
$0.5 < m < 1$ with a non-vanishing but finite autocorrelation time
$\Dt<\tau<\infty$. Here, the network maintains and amplifies input as
recurrently generated fluctuations. In these two regimes the mean-field
solution \eqref{eqMFprediction} matches numerical data on different network
topologies (Fig.~\ref{figHomPlastDynamicState}). Third, the mean-field solution
predicts that in the limit $h\to 0$ the dynamics become critical with divergent
autocorrelation time ($m\to1$, $\tau\to\infty$). However, we observe a clear
deviation from the mean-field solution, which defines the \textit{bursting
regime} with $m>1$ and a finite autocorrelation time, as discussed below. 

\subsection{Bursting regime}
Deviations from the mean-field solution \eqref{eqMFprediction} emerge when the
assumption of ``sufficiently large $\tauhp$'' breaks down. We will derive a
bound for $\tauhp$, below which the (rapid) homeostatic feedback causes  notable
changes of the network branching parameter $\avg{m}_t$ around its mean
$m=\expectation{\avg{m}}$, which in turn jeopardize the stability of the network
dynamics. 

To estimate the change of the network branching parameter, we first consider the
change in local branching parameter $\Dm_{i,t}$, which depends on each neurons
out-degree $k_i = \sum_{j=1}^N w_{ij}$ and is given by
\begin{equation*}\label{eqHomeostaticPlasticityLocal}
  \Dm_{i,t} = \sum_{j=1}^N w_{ij} \Dalpha_{j,t} = \left(k_i\Dt\,r\target-\sum_{j=1}^{N}
  w_{ij} \, s_{j,t}\right)\left(\frac{\Dt}{\tauhp}\right).
\end{equation*}
On the network level, we make the assumption that the state of each neuron is
approximated by the network average $s_{i,t}\approx A_t/N$, such that
$\avg{\sum_{j=1}^N w_{ij} \, s_{j,t}}\approx \avg{k} \frac{A_t}{N}$. Then, the
change in network branching parameter can be approximated as 
\begin{align}\label{eqHomeostaticPlasticityNetwork}
  \Delta\avg{m}_t = \avg{\Dm}_t
  &\approx\left(\avg{k}\Dt\,r\target-A_t\frac{\avg{k}}{N}\right)\left(\frac{\Dt}{\tauhp}\right)\nonumber\\
  &\approx\left(\Dt\,r\target-\frac{A_t}{N}\right)\left(\frac{\Dt}{\tauprime}\right),
\end{align}
where we have introduced an effective homeostatic network timescale
$\tauprime=\tauhp/\avg{k}$, for which \eqref{eqHomeostaticPlasticityNetwork}
recovers the form of \eqref{eqHomeostaticPlasticityNeuron}. 
Using $\tauprime$ allows one to semi-analytically approximate the deviation of
$m$ from the mean-field solution (Fig.~\ref{figHomPlastDynamicState}\textbf{a},
dashed lines, and Appendix~\ref{secAppendixNonStat}).

We next show that the stability of network dynamics requires the autocorrelation
time of the dynamic process $\tau$ to be smaller than the timescale of
homeostasis $\tauprime$. Stability demands that the homeostatic change in
autocorrelation time $\Delta\tau$ is small compared to the autocorrelation time
itself, i.e., $\Delta\tau\ll\tau$. We approximate $\Delta\tau$ by error
propagation in \eqref{eqTau}, yielding
\begin{equation}\label{eqDeltaTau}
  \Delta \tau \simeq \left|(\tau^2/\Dt)
  e^{\Dt/\tau}\right|\Dm\simeq(\tau^2/\Dt+\tau)\Dm,
\end{equation}
where we expanded the exponential for small $\Dt/\tau$.
For large $\tau$, the leading term in \eqref{eqDeltaTau} dominates and
inserting \eqref{eqHomeostaticPlasticityNetwork} yields
%
  $\Delta\tau\simeq\left|\Dt\,r\target-A_t/N\right|\left(\tau^2/\tauprime\right)$.
%
Thus, the dynamics can be described as a stationary
branching process (mean-field solution) only as long as  
\begin{equation}\label{eqBurstingOnset}
  \tau \ll \tau^\prime \left|\Dt~r^\ast-A_t/N\right|^{-1}.
\end{equation}
Violation of \eqref{eqBurstingOnset} results in bursting behavior
(Figs.~\ref{figOverviewMF}\,\&\,\ref{figNonStat}). 
For $A_t=\mathcal{O}(N)$ the right hand side of \eqref{eqBurstingOnset} is
minimal, because  $\Dt\, r\target\ll1$, which implies a maximal attainable
autocorrelation time $\tau\simeq\tauprime=\tauhp/\overline{k}$. This is in
perfect agreement with the saturation of measured autocorrelation time in the
bursting regime (Fig.~\ref{figHomPlastDynamicState}\textbf{b}, dashed lines).

The transition from the fluctuating to the bursting regime occurs when the
mean-field solution \eqref{eqMFprediction} equals the maximal attainable
autocorrelation time, i.e., $\tau=-\Delta t/\ln(1-h/r^\ast)\approx\tauprime$.
Hence, the transition occurs at $h/r\target\approx
1-e^{-\Dt/\tauprime}\approx\Dt/\tauprime$. For even lower input rate, the
dynamics become more and more bursty, and the avalanche-size distribution
exhibits a peak at large avalanche sizes (Fig.~\ref{figOverviewMF} for
$h/r\target<10^{-2}$, where $\tauprime=\SI{e2}{ms}$, $\Dt=\SI{1}{ms}$). At the
transition, the dynamics can be considered close-to-critical, because the (fully
sampled) avalanche-size distribution is closest to a power-law with exponent
$-3/2$.


\subsection{Distributions of spiking activity}
\begin{figure}
  \includegraphics{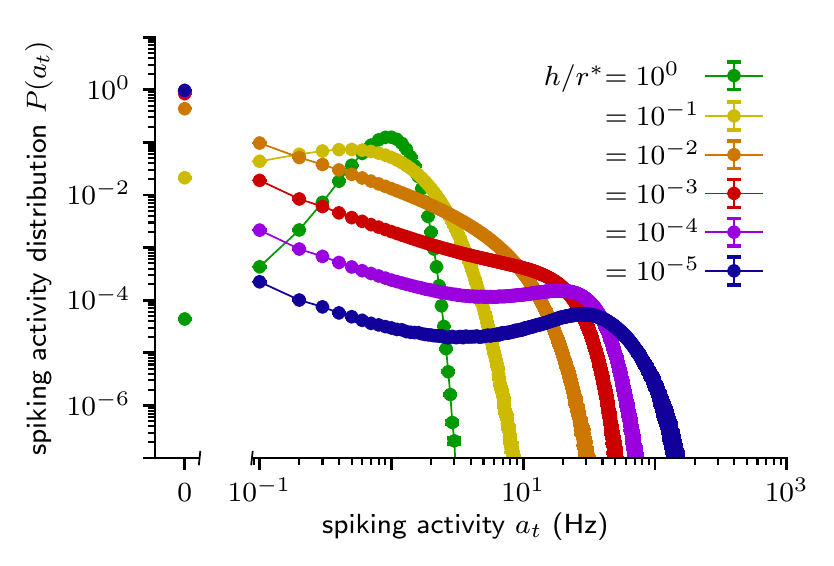}
  \caption{%
    Distribution of spiking activity in weakly connected Erd\H{o}s-R{\'e}nyi
    networks ($\pER=10^{-2}$, $\Dt=\SI{1}{ms}$, $r\target=\SI{1}{Hz}$,
    $\tauprime=\SI{e4}{ms}$) averaged over 12 independent simulations. For
    irregular dynamics ($h/r\target\approx10^0$) the distribution is clearly
    unimodal. For fluctuating dynamics ($10^{-4}<h/r\target<10^0$) the
    distribution broadens and shifts the maximum towards quiescence. In addition
    towards the lower bound of the regime, the distribution develops a power-law
    tail with an exponential cutoff. At the crossover to bursting dynamics
    ($h/r\target\approx10^{-4}$) the distribution becomes bimodal.
    \label{figUpDown}
  }
\end{figure}

The different dynamical regimes imply characteristic distributions of neural
network activity $P(a_t)$.  
Figure~\ref{figUpDown} shows an example of $P(a_t)$ for ER networks with
$\pER=10^{-2}$, where the transition from fluctuating to bursting dynamics is
expected for $h/r\target\approx\Dt/\tauprime=10^{-4}$. In the irregular regime
(green) $P(a_t)$ is a unimodal distribution. In the fluctuating regime
(yellow\--red), the peak in $P(a_t)$ shifts towards quiescence and the
distribution develops a power-law tail with exponential cutoff, expected for a
critical branching process. In the bursting regime (purple\--blue), $P(a_t)$ is
a bimodal distribution, reflecting network changes between quiescence and bursty
activity. The position and sharpness of the high-activity maximum depend on the
network connectivity and hence the heterogeneity in the single-neuron input. 

\subsection{Reproducing experimental results}
\label{secResultsMatch}
Using the insight from our theory, we can reproduce experimental results.
Spiking activity recorded \invivo{} resembles dynamics of the fluctuating
regime. In this regime, the dynamic state is consistent for all topologies we
considered (Fig.~\ref{figHomPlastDynamicState}). Therefore, already a branching
network on an AA topology suffices to quantitatively reproduce the
avalanche-size distributions by matching model parameters with experimentally
accessible estimates (Fig.~\ref{figExperiment} dashed lines). To match the
branching network to recordings from cat V1 and rat CA1, we first estimated the
spike rate $r$ and autocorrelation time $\tau$ from the recordings of spiking
activity~\cite{wilting2018}; we then chose biologically plausible parameters for
the network size $N$, the homeostatic timescale $\tauhp$, as well as the
simulation time step $\Dt$; and finally derived the external input $h$ using
\eqref{eqMFprediction} (for details see Appendix~\ref{secAppendixMatchExp}).
The resulting subsampled avalanche-size distributions are in astonishing
agreement with the experimental results, given the simplicity of our approach.
Close inspection of the avalanche-size distribution for rat CA1 recordings still
reveals small deviations from our model results. The deviations can be
attributed to theta-oscillations in hippocampus, which result in subleading
oscillations on an exponentially decaying autocorrelation
function~\cite{wilting2018}. While this justifies our approach to consider a
single dominant autocorrelation time, theta oscillations slightly decorrelate
the activity at short times and thereby foster premature termination of
avalanches. Thus, the tail in the avalanche-size distribution is slightly
shifted to smaller avalanche sizes (Fig.~\ref{figExperiment}).

The \invitro{} results are qualitatively well matched by simulations in the
bursting regime, with avalanche-size distributions showing a characteristic peak
at large avalanche sizes (Fig.~\ref{figOverviewMF}). It is difficult to
quantitatively match a model to the data, because a number of parameters can
only be assessed imprecisely. Most importantly, the autocorrelation time in the
burst regime is not informative about the external input rate $h$ and depends on
the average number of connections (Fig.~\ref{figHomPlastDynamicState}).
Likewise, the time-dependence of the branching parameter $m_t$ cannot be
assessed directly.  Finally, system size and topology impact the network
dynamics more strongly in this regime than in the fluctuating or input-driven
regime. This yields a family of avalanche-size distributions with similar
qualitative characteristics but differences in precise location and shape of the
peak at large sizes. 
\section{Discussion}

\begin{figure}[]
  \includegraphics{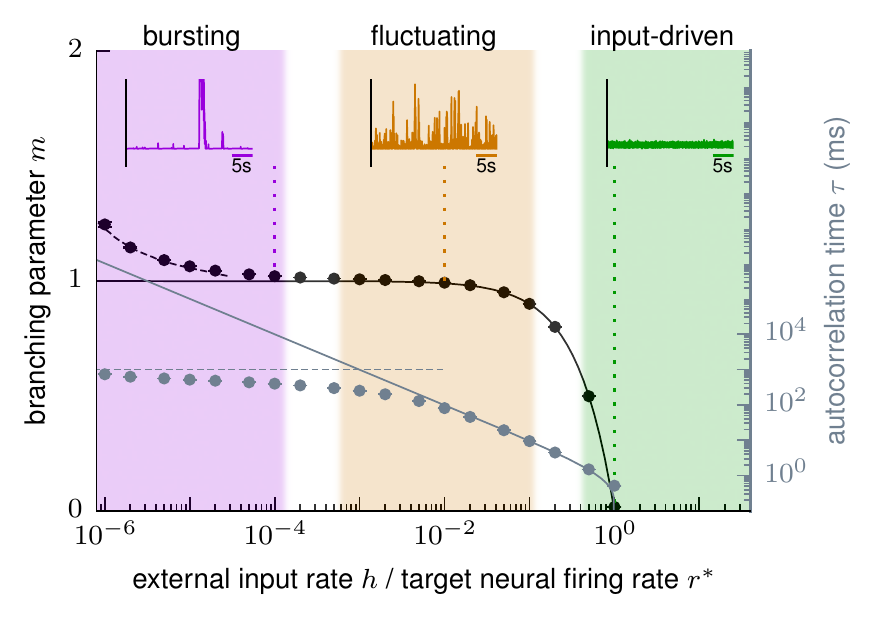}
  \caption{%
    Sketch of bursting, fluctuating and input-driven network states, classified
    by the branching parameter and the autocorrelation time. We propose (solid
    lines) that homeostatic plasticity tunes the dynamic state depending on the
    ratio of external input rate (including spontaneous neural firing) and
    target neural firing rate.
    Data points and example activity traces stem from Erd\H{o}s-R{\'e}nyi
    networks ($N=10^4$, $p=10^{-1}$, $\tauhp=\SI{e3}{s}$).
    In the bursting regime, the homeostatic timescale $\tauhp$ influences the
    resulting dynamics (dashed lines). 
    \label{figSketch}
    \vspace{-0.5em}
  }
\end{figure}

We propose the interplay of external input rate and target spike rate, mediated
by homeostatic plasticity, as a neural mechanism for self-organization into
different dynamic states (cf. sketch in Fig.~\ref{figSketch}). Using the
framework of a branching process, we disentangled the recurrent network dynamics
from the external input (e.g. input from other brain areas, external stimuli and
spontaneous spiking of individual neurons). Our mean-field solutions,
complemented by numeric results for generic spiking neural networks, show that
for high input the network organizes into an input-driven state, while for
decreasing input the recurrent interactions are strengthened, leading to a
regime of fluctuating dynamics, resembling the reverberating dynamics observed
\invivo{}.  Decreasing the input further induces bursting behavior, known from
\invitro{} recordings, due to a competition of timescales between homeostatic
plasticity and the autocorrelation of population activity. Thereby our framework
proposes a generic mechanism to explain the prominent differences between
\invivo{} and \invitro{} dynamics.


Our theory suggests that also differences within the collective dynamic state
observed \invivo{} can be explained by considering differences in input
strength. 
For cortex, it was shown that layer 2/3 exhibits critical
dynamics~\cite{bellay2015} and presumably deeper layers show 
reverberating dynamics~\cite{priesemann2014, wilting2018}. We propose that this can be
caused by different input strength: layer 2/3 is more recurrently connected,
while layer 4 is the main target of thalamic input~\cite{harris2013}, hence
receiving the stronger input. 
The dynamic state varies also across different cortical areas, where
autocorrelation times of network activity reflect a hierarchical
organization~\cite{murray2014,hasson2015}: Cortical areas associated with higher-order function
show a larger autocorrelation time. In the light of our results, a larger
autocorrelation time implies less afferent input for the area in question.
The hierarchical organization is further supported by our analysis of spiking
activity \invivo{} (Fig.~\ref{figExperiment}): the autocorrelation times in
visual cortex ($\tau\approx\SI{0.2}{s}$) and hippocampus
($\tau\approx\SI{2}{s}$) precisely reflect that visual cortex is almost at
the bottom, whereas hippocampus is at the
top of the hierarchy of visual processing~\cite{felleman1991}.

Our theory provides an approach to experimentally infer the fraction of spikes
generated recurrently within a network and generated by external input. For an
average spike rate $r$, equation \eqref{eqMFprediction} implies
$h/r=(1-e^{-\Dt/\tau})$. The external input rate can then be directly calculated
from the autocorrelation time and by assuming a biologically plausible
signal-propagation time, e.g., $\Dt\approx\SI{4}{ms}$.
%
We estimate for recordings from visual cortex in mildly anesthetized cat that
about $2\%$ of the network activity is generated by the input, whereas the
majority of $98\%$ are generated recurrently within the network. 
From autocorrelation times measured across the cortical hierarchy ($\SI{50}{ms}$
to $\SI{350}{ms}$) in macaque monkey~\cite{murray2014}, the fraction of spikes
generated by external input  decreases from $\sim\SI{8}{\%}$ to $\sim\SI{1}{\%}$
from lower to higher cortical areas. 
This is consistent with perturbation experiments in rat barrel cortex, where
after triggering an extra spike the decay time of population rate was at least
$\SI{50}{ms}$~\cite{london2010} indicating at most about $8\%$ external input
(for a detailed discussion see also Ref.~\cite{wilting2018b}).
%
Last, experiments on visual cortex of awake mice directly after thalamic silencing
found a decay time of $\tau=\SI{12(1)}{ms}$~\cite{reinhold2015}, from which we
would estimate about $70\%$ recurrent activation. This is in perfect agreement
with the experimentally measured $72(6)\%$ of recurrent activation in the same
study. This result thus validates our derived relation between $h/r$ and $\tau$. 
One can interpret our findings in the light of up and down
states~\cite{wilson2008, stern1997, holcman2006, millman2010}. Because the
membrane potential was found to correlate with network
activity~\cite{cossart2003, vardi2016}, our results for the distribution of
spiking activity in the bursting regime may correspond to the bimodal
distributions of membrane potentials during up and down states
(Fig~\ref{figUpDown}). It has already been shown that negative feedback can
stabilize up and down states~\cite{holcman2006, millman2010}. In our theory,
negative feedback leads to similar results in the low-input regime. Moreover, we
predict that decreasing network input further, prolongs the quiescent periods or
down states.


Our theory unifies previous numerical approaches of self-organization in neural
networks, which typically considered a negative feedback mechanism but made 
very different choices on a (fixed) network input.
For example, bursting dynamics have been generated by homeostatic build-up upon
loss of network input~\cite{froehlich2008a} or by self-organized
supercriticality through dynamic neuronal gain~\cite{costa2017}. 
Adding weak input, self-organized criticality~\cite{bak1987,zapperi1995} has
been achieved by local rewiring~\cite{bornholdt2000, tetzlaff2010, kossio2018} and synaptic
depression~\cite{arcangelis2006, levina2007, levina2009, bonachela2010,
costa2015, kessenich2016, campos2017, herrmann2017}.
In contrast, asynchronous-irregular network activity typically requires
comparably strong input, assuming a balanced state~\cite{vreeswijk1996,
brunel2000, renart2010}, and a self-organized AI network state can be promoted
by inhibitory plasticity~\cite{vogels2011,effenberger2015}. 
While all these studies provide mechanisms of self-organization to one
particular dynamic state, our theory highlights the role of input in combination
with a negative feedback~\cite{naude2013,brunel2000,lercher2015,munoz2018} and
provides a unifying mechanism of self-organization covering bursting,
fluctuating and irregular dynamics. 

From a broader perspective, we characterized driven systems with a negative
feedback as a function of the input rate. The negative feedback compensates the
input by regulating the system's self-activation to achieve a target activity.
%
In this light of control theory, the bursting regime can be understood as
resonances in a feedback loop, where feedback dynamics are faster than system
dynamics (cf.~\cite{harnack2015}). 
This qualitative picture should remain valid for other connected graphs subject
to external input with spatial and temporal correlations. In this case, however,
we expect more complex network responses than predicted by our mean-field
theory, which assumes self-averaging random networks subject to uncorrelated
input.

Our results suggest that homeostatic plasticity may be exploited in experiments
to generate \invivo{}-like dynamics in a controlled \invitro{} setup, in
particular to abolish the ubiquitous bursts \invitro{}. Previous attempts to
reduce bursting \invitro{}~\cite{wagenaar2005} and in model-systems of
epilepsy~\cite{lian2003, chiang2014, covolan2014, ladas2015} used short-term
electrical and optical stimulation to attain temporal reduction in bursting.
Alternatively, one can reduce bursting pharmacologically or by changing the
calcium level, however, typically at the cost of changing single-neuron
properties~\cite{morefield2000, shew2009, penn2016}.
We propose a different approach, namely applying weak, global, long-term
stimulation. Mediated by homeostasis, the stimulation should alter the
effective synaptic strength, and thereby the dynamic state while preserving
single-neuron dynamics~\cite{schottdorf2017}.
%
%
In particular, we predict that inducing in every neuron additional spikes with
$h=\mathcal{O}(\SI{0.01}{Hz})$ is sufficient to abolish the ubiquitous bursts
\invitro{} and render the dynamics \invivo{}-like instead.
If verified, this approach promises completely novel paths for drug studies. By
establishing \invivo{}-like dynamics \invitro{}, fine differences between
neurological disorders, which are otherwise masked by the ubiquitous bursts, can
be readily identified. Altogether this would present a comparably
cost-efficient, high-throughput, and well-accessible drug assay with largely
increased sensitivity.

\begin{acknowledgments}
  We would like to thank Manuel Schottdorf and Andreas Neef, as well as
  Raoul-Martin Memmesheimer and Sven Goedeke, for stimulating discussions. We
  are grateful for careful proofreading from Roman Engelhardt, Jo{\~ a}o
  Pinheiro Neto, and Conor Heins. All authors acknowledge funding by the Max
  Planck Society. JZ and VP received financial support from the German Ministry
  of Education and Research (BMBF) via the Bernstein Center for Computational
  Neuroscience (BCCN) G{\"o}ttingen under Grant No.~01GQ1005B.  JW was
  financially supported by Gertrud-Reemtsma-Stiftung and Physics-to-Medicine
  Initiative G{\"o}ttingen (ZN3108) LM des Nieders{\"a}chsischen Vorab. 
\end{acknowledgments}

\begin{appendix}
  \renewcommand{\theparagraph}{\arabic{paragraph}}

  \section{Experimental details }
  \label{secAppendixExperiments}
  \paragraph{Dissociated dense cultures of cortical rat neurons:}
  \label{secAppendixWagenaar}
  %
  The spike-time data from dissociated cortical rat neurons of mature dense
  cultures was recorded by Wagenaar \textit{et al.}~\cite{wagenaar2006} and was
  obtained freely online~\cite{wagenaarURL}. The experimental setup uses
  multi-electrode-arrays (MEA) with $n=59$ electrodes. Cortical cells were
  obtained from dissecting the anterior part of the cortex of Wistar rat embryos
  (E18), including somatosensory, motor, and association areas.  For details, we
  refer to \cite{wagenaar2006}. Measurements were performed every day \invitro{}
  (DIV). We here focus on the dense case with $50\,000$ cells plated initially
  with a density of $2.5(1.5)\times10^3$ cells/\si{mm$^2$} at 1~DIV, which is
  compatible with standard \invitro{} experiments in the field that claim to
  observe critical dynamic behavior. We selected the representative recordings
  8-2-34 (exp 1) and 7-2-35 (exp 2) at mature age (34/35 DIVs) for
  Fig.~\ref{figExperiment}.
  
  \paragraph{Rat hippocampus:}
  \label{secAppendixRat}
  The spiking data from rats were recorded by Mizuseki \textit{et
  al.}~\cite{mizuseki2009a, mizuseki2009b} with experimental protocols approved
  by the Institutional Animal Care and Use Committee of Rutgers University. The
  data was obtained from the NSF-founded CRCNS data sharing
  website~\cite{mizuseki2009b}. The spikes were recorded in CA1 of the right
  dorsal hippocampus during an open field task. Specifically, we used the data
  set ec013.527 with sorted spikes from 4 shanks with $n=31$ channels. For details
  we refer to Refs.~\cite{mizuseki2009a, mizuseki2009b}.
  
  \paragraph{Primary visual cat cortex:}
  \label{secAppendixBlanche}
  The spiking data from cats were recorded by Tim Blanche in the laboratory of
  Nicholas Swindale, University of British Columbia, in accordance with
  guidelines established by the Canadian Council for Animal
  Care~\cite{blanche2006, blanche2009}. The data was obtained from the
  NSF-founded CRCNS data sharing website~\cite{blanche2009}. Specifically, we
  used the data set pvc3 with recordings of $n=50$ sorted single
  units~\cite{blanche2006} in area 18. For details we refer to
  Refs.~\cite{blanche2009, blanche2006}. We confined ourselves to the
  experiments where no stimuli were presented such that spikes reflect the
  spontaneous activity in the visual cortex of mildly anesthetized cats. 
  In order to circumvent potential non-stationarities at the beginning and end
  of the recording, we omitted the initial \SI{25}{s} and stopped after
  \SI{320}{s} of recording~\cite{wilting2018}.

  \section{Analysis details}
  \label{secAppendixNumerics}
  \paragraph{Spiking activity: }
  \label{secAppendixNeuralActivity} In order to present the spiking activity over
  time, we partition the time axis of experimental or numerical data into
  discrete bins of size $\Dt$. For the time-discrete simulations the time bin
  naturally matches the time step. For experimental data we set
  $\Dt=\SI{4}{ms}$.  In each time bin we count the total number of spikes $A_t$
  and normalize with the number of neurons $N$ to obtain the average spiking
  activity $a_t=A_t/N\Dt$. Note that experimental preparations were inevitably
  subsampled, as spikes were recorded only from a small number of all neurons.

  \paragraph{Avalanche-size distribution: }
  \label{secAppendixAvalanche}
  We define the avalanche size $s$ as the number of spikes enclosed along the
  discrete time axis by bins with zero activity~\cite{beggs2003}. To test for
  criticality in terms of a branching process, one compares $P(s)$ to the
  expected $P(s)\sim s^{-3/2}$.  This is a valid approach in the limit
  $h\rightarrow 0$, where avalanches can be clearly identified, and for fully
  sampled systems~\cite{priesemann2014}.  However, experiments are limited to
  record only from $n$ out of $N$ neurons.  As a result, the distributions for
  subsampled activity $P_{\rm sub}(s)$ differ due to subsampling
  bias~\cite{priesemann2009, priesemann2013}. 
  Therefore, we numerically measure both full ($n=N$) and subsampled ($n<N$)
  avalanche-size distributions to qualitatively compare $P(s)$ to the theory and
  $P_{\rm sub}(s)$ to experimental data.

  \paragraph{Integrated autocorrelation time: }
  \label{secAppendixTau}
  We measure the autocorrelation time of spiking activity $a_t$ in terms of the
  integrated autocorrelation time $\tau_\mathrm{int}$, for details see, e.g.,
  Ref.~\cite{janke2002}. In brief, we sum over the normalized autocorrelation
  function $C(l)=\mathrm{Cov}[a_t, a_{t+l}]/\mathrm{Var}[a_t]$ until the sum converges. Following
  conventions, we define
  $\tau_\mathrm{int}=\Dt[\frac{1}{2}+\sum_{l=1}^{l_\mathrm{max}} C(l)]$, where
  $l_\mathrm{max}$ is self-consistently obtained as the minimal $l_\mathrm{max}
  > 6\,\tau_\mathrm{int}(l_\mathrm{max})$. 

  
  \paragraph{Reproducing experimental results:}
  \label{secAppendixMatchExp}
  We use a branching network with AA topology subject to homeostatic
  plasticity to quantitatively reproduce \invivo{} subsampled avalanche-size
  distributions. We chose networks of size $N=10^4$ with sufficiently large
  homeostatic timescale $\tau_\mathrm{hp}=\SI{e5}{s}$.
  The following model parameters can be obtained from experimentally measured
  values: In the chosen recordings, we measured the average rate
  ($r_\mathrm{cat}\approx\SI{7}{Hz}$ and $r_\mathrm{rat}\approx\SI{11}{Hz}$) as
  well as the subsampling corrected branching parameter~\cite{wilting2018}
  ($m_\mathrm{cat}\approx0.98$ and $m_\mathrm{rat}\approx 0.997$ for
  $\Dt=\SI{4}{ms}$). In fact, the branching parameter is not suitable to
  identify the input rate via \eqref{eqMFprediction}, because it refers to a
  process in discrete time steps. Since we are treating a continuous process,
  the invariant quantity is the autocorrelation time
  ($\tau_\mathrm{cat}\approx\SI{0.2}{s}$ and
  $\tau_\mathrm{rat}\approx\SI{1.6}{s}$). According to our theory, we can then
  calculate the input rate per neuron $h=(1-\exp(-\Dt/\tau))r$. In order to
  avoid convergence effects, we need to choose a sufficiently small time step
  $\Dt=\SI{1}{ms}$ of signal propagation (resulting in
  $h_\mathrm{cat}\approx\SI{3.5e-2}{Hz}$ and
  $h_\mathrm{rat}\approx\SI{5.5e-3}{Hz}$), while we record in time bins of
  $\SI{4}{ms}$ to match the analysis of the experiments. Subsampled
  avalanche-size distributions are estimated by randomly choosing $n<N$ neurons,
  where we approximated $n$ by the number of electrodes or channels
  ($n_\mathrm{cat}=50$ and $n_\mathrm{rat}=31$).
  
  \section{Approximating the dynamic state in the bursting regime}
  \label{secAppendixNonStat}
  \begin{figure}
    \includegraphics{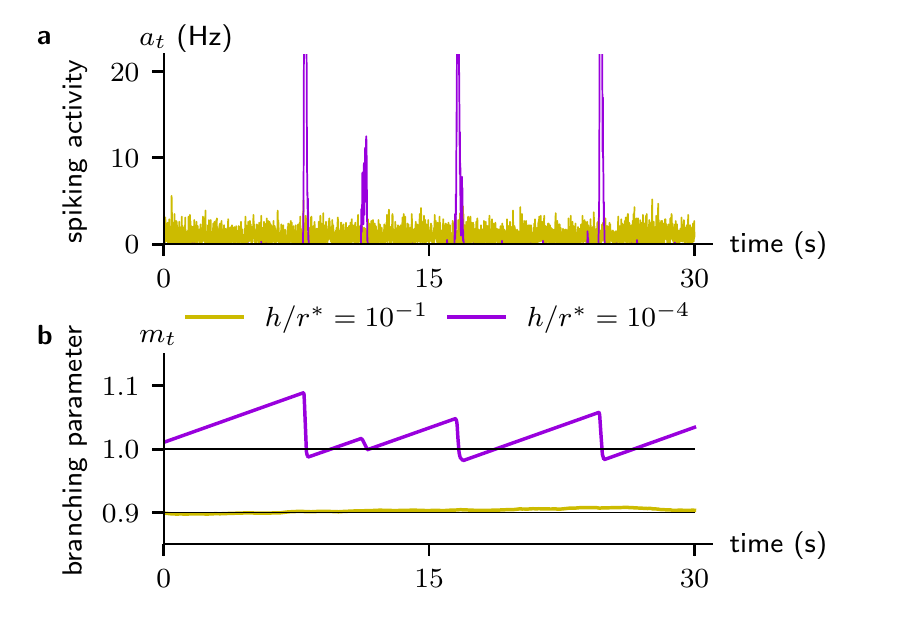}
    \vspace{-0.5cm}
    \caption{%
      Temporal fluctuations in an annealed-average network with homeostatic plasticity
      subject to different external input rates.
      (\textbf{a}) Spiking activity $a_t$ shows small fluctuations for large
      input rates (yellow) and bursts for small input rates (purple), cf.
      Fig.~\ref{figOverviewMF}. (\textbf{b}) Branching parameter
      fluctuates around predicted value (black horizontal lines) and develops
      distinct saw-tooth pattern for small input rates.
      \label{figNonStat}
      }
  \end{figure}
  We showed in Sec.~\ref{secResults} that decreasing the external input to 
  recurrent networks with homeostatic plasticity leads to bursting behavior
  (Fig.~\ref{figNonStat}\textbf{a}). This is directly related to the network
  branching parameter $m_t=\avg{m}_t$ no longer showing small fluctuations
  around the predicted value but instead exhibiting a prominent saw-tooth
  pattern (Fig.~\ref{figNonStat}\textbf{b}), a hallmark of the homeostatic
  buildup in the long pauses with no input. 

  We here show a semi-analytical approximation of the network branching
  parameter in the bursting regime. For sufficiently small external input we may
  assume separation of timescales, i.e., every externally induced spike drives
  one avalanche with periods of silence in between. Let us first consider the
  periods of silence, i.e., no activity per site. This holds during the entire
  growth period $T$ such that \eqref{eqHomeostaticPlasticityNetwork} yields 
  \begin{equation}\label{eqNonStatDevelop}
    m_t-m_{t-T} = (\Dt r\target)\frac{T}{\tauprime}.
  \end{equation}
  The situation becomes more involved within the bursts, where the behavior of
  $m_t$ is nonlinear. Consider an external spike that triggers an avalanche at
  $t=s$ which ends at $t=e$. Due to the separation of timescales we can assume
  $A_s=1$. There are two possible scenarios: (i) The avalanche dies out before a
  burst can develop and (ii) the input triggers a proper burst with a macroscopic
  activation.  
  
  We first estimate the probability that an avalanche dies out before a burst
  develops. For $\tauhp\gg\Dt$ we approximate $m_t\approx m_s=\text{const}$.
  Then, the probability of ultimate extinction $\theta$ can be calculated as the
  solution of $\theta=\Pi(\theta)$ with $\Pi(\theta)$ the probability generating
  function~\cite{harris1963}. In the onset phase, the branching process is
  described by a Poisson process per event with mean $m_s$, such that
  $\Pi(\theta)=e^{-m_s(1-\theta)}$. We are thus looking for a solution
  of 
  \begin{eqnarray}
    \theta = e^{-m_s(1-\theta)}, 
  \end{eqnarray}
  which can be rewritten to 
  \begin{eqnarray}
    -m_s\theta e^{-m_s\theta}=-m_se^{-m_s}.
  \end{eqnarray}
  We identify the Lambert-W function $W(z)e^{W(z)}=z$~\cite{corless1996} with
  $W(z)=-m_s\theta=W(-m_se^{-m_s})$ and find for the probability that no
  burst develops
  \begin{eqnarray}\label{eqNonStatProbabilityBurst}
    p_{\rm no-burst}(m_s)= \theta = -\frac{1}{m_s} W\left(-m_s e^{-m_s}\right).
  \end{eqnarray}

  If a proper burst develops, the strong activity diminishes $m_t$ until the
  burst dies out again. We cannot analytically estimate the branching parameter
  $m_e$ after burst end, but we can use a deterministic numerical approximation
  to obtain $m_e(m_s)$. Instead of stochastically generating new (discrete)
  events according to some distribution $P(m_t)$ with average $m_t$, we
  approximate the branching process as deterministic (continuous) evolution
  $A_{t+1}=m_t A_t$. For a finite network, we need to consider convergence
  effects when one neuron is activated by two or more neurons at the same time.
  In the absence of external input, this introduces for an AA (i.e.
  approximating fully connected) network the activity-dependent branching
  parameter~\cite{levina2018} 
  \begin{equation}
    m_t(A_t)=\frac{N}{A_t}\left(1-\left(1-\frac{m_t}{N}\right)^{A_t}\right),
  \end{equation}
  which we need to consider for the activity propagation within the burst, i.e.,
  $A_{t+1}=m_t(A_t)A_t$. In addition, we introduce an upper bound $A_{t}\leq N$.
  The upper limit on $A_t$ puts a lower bound on $\Dm_t$ according to
  \eqref{eqHomeostaticPlasticityNetwork} and thus extends the duration of
  avalanches. Evolving $m_{t+1}=m_t+(\Dt r\target - A_t/N)(\Dt/\tauprime)$, with
  $m_{t+1}\geq0$, we iterate until $A_e<1$.  This is a quick and
  numerically robust iterative scheme to estimate $m_e(m_s)$.

  Putting everything together, we numerically approximate the average network
  branching parameter $m$ under homeostatic plasticity in the bursting regime of
  low external input for an AA network. For this, we sample the external spikes
  (drive) as $10^4$ inter-drive intervals $T_s$ from an exponential distribution
  $P(T)=(1/hN)e^{-T/hN}$, corresponding to $N$ Poisson processes with rate $h$.
  The remaining part can be interpreted as an event-based sampling with
  approximate transformations: Starting with $m_0=0$, we evolve $m_t$ for each
  inter-drive interval $T_s$ according to \eqref{eqNonStatDevelop}. If $m_t>1$,
  we keep $m_t$ with probability $p_{\rm no-burst}(m_t)$ or else initiate a
  burst by setting $m_t=m_e\left(m_t\right)$.  Afterwards we continue evolving
  $m_t$. 

  \section{Characteristic duration of inter-burst-intervals in burst regime}
  \label{secAppendixPeriodicity}
  \begin{figure}
    \includegraphics{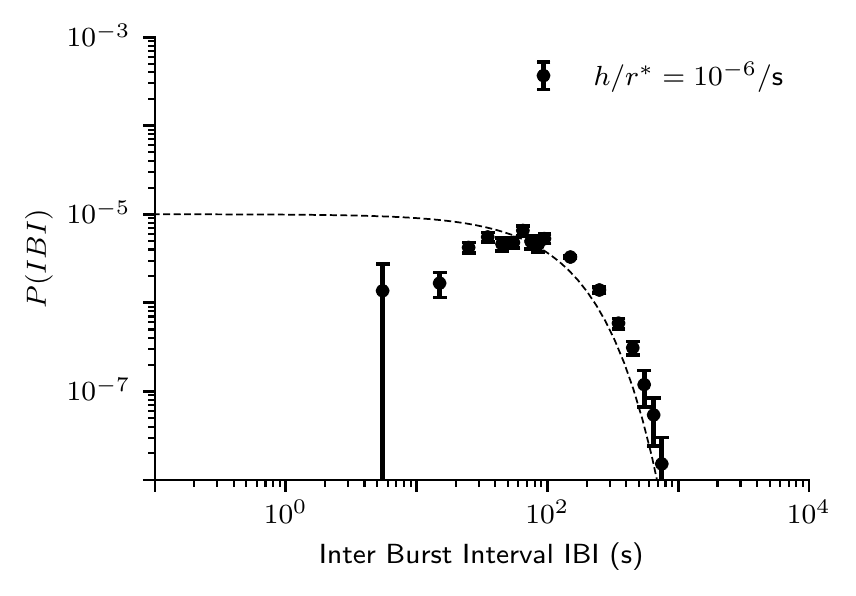}
    \caption{%
      Inter-burst-interval (IBI) distribution for annealed-average networks
      averaged over 12 independent simulations. Intervals are measured as times
      between proper burst onsets ($a_t>20r\target$). Dashed lines show the
      exponential inter-spike-interval distribution of the Poisson external
      drive.
      \label{figDistIBI}
      }
    \end{figure}
    In the bursting regime of low external input, the spiking activity suggests
    a characteristic time between bursts. In order to test for periodicity, we
    analyzed the distribution of inter-burst-intervals (IBI), where intervals
    are measured as the time between two consecutive burst onsets, defined as a
    spiking activity $a_t > 20 r\target$. We find (Fig.~\ref{figDistIBI}) that
    large IBI are suppressed by the exponentially distributed inter-drive
    intervals (dashed lines), while short IBI are suppressed by the probability
    $p_\mathrm{no-burst}(m)$ that a given external spike does not trigger a proper
    burst (Appendix~\ref{secAppendixNonStat}). This gives rise to a
    characteristic duration of inter-burst-intervals in the burst regime,
    although the dynamics are not strictly periodic.

\end{appendix}


\end{document}